\documentclass[nofootinbib,notitlepage,11pt]{revtex4-2}
\pdfoutput=1

\usepackage[dvipsnames]{xcolor}
\usepackage{hyperref}
\hypersetup{
  breaklinks=true,
  colorlinks=true,
  allcolors=BlueViolet,
}

\usepackage[yyyymmdd]{datetime} % set date format
\usepackage[inline]{enumitem}  % inline enumeration
\setlist{nosep} % more compact spacing between environments

\usepackage{physics}  % basic math
\usepackage{braket}  % braket notation
\usepackage{amssymb}  % for \mathbb
\newcommand{\p}[1]{\left(#1\right)} % parenthesis
\renewcommand{\sp}[1]{\left[#1\right]} % square parenthesis
\renewcommand{\set}[1]{\left\{#1\right\}} % curly parenthesis
\newcommand{\bk}{\braket} % shorthand for braket notation

\def\obra#1{\mathinner{({#1}|}}
\def\oket#1{\mathinner{|{#1})}}

\def\oop#1#2{\oket{#1}\!\obra{#2}}

\newcommand{\B}{\mathcal{B}}
\newcommand{\C}{\mathcal{C}}
\newcommand{\E}{\mathcal{E}}
\renewcommand{\H}{\mathcal{H}}
\renewcommand{\O}{\mathcal{O}}
\newcommand{\U}{\mathcal{U}}

\newcommand{\EE}{\mathbb{E}}

\newcommand{\ent}{\mathrm{ent}}
\newcommand{\err}{\mathrm{err}}
\newcommand{\Pauli}{\mathrm{Pauli}}
\newcommand{\ii}{\mathrm{i}\mkern1mu} % imaginary unit

\newtheorem{theorem}{Theorem}
\newtheorem{corollary}{Corollary}

\let\dim\relax
\DeclareMathOperator{\dim}{dim}

\usepackage{tikz}
\usetikzlibrary{calc}

\begin{document}
\count\footins = 1000 % allows for longer footnotes
\sloppy % prevent text from spilling into the margins

\title{A short note on effective Pauli noise models}%
\author{Michael A. Perlin}%
\email{mika.perlin@gmail.com}%
\affiliation{Infleqtion, Inc., Chicago, IL, 60615, USA}%
% \date{\today}

\begin{abstract}
  We provide a simple prescription to extract an effective Pauli noise model from classical simulations of a noisy experimental protocol for a unitary gate.
  This prescription yields the closest Pauli channel approximation to the error channel associated with the gate implementation, as measured by the Frobenius distance between quantum channels.
  Informed by these results, we highlight some puzzles regarding the quantitative treatment of coherent errors.
\end{abstract}

\maketitle

%%%%%%%%%%%%%%%%%%%%%%%%%%%%%%%%%%%%%%%%%%%%%%%%%%
\section{Introduction}

Noise models are important for benchmarking quantum computers as they scale up to meet the requirements for quantum error correction (QEC).
While a physics-informed analysis can provide a comprehensive picture of errors in a system with a handful of qubits, a detailed analysis rapidly becomes intractable with growing system size.
For this reason, large scale error analyses are typically performed with simplified noise models, such as those assuming only Pauli errors.

Motivated by theoretical proposals for novel quantum gate protocols such as Refs.~\cite{saffman2020symmetric, young2021asymmetric, farouk2022parallel} and their potential use in QEC codes \cite{crow2016improved, perlin2023faulttolerant}, in this note we work out how to extract an effective Pauli noise model from classical physics-level simulations of a quantum gate protocol.
Specifically, we consider a target $N$-qubit unitary $U_0$, such as a \texttt{CZ} gate, and an approximate implementation $U\approx U_0$.
We are interested in the scenario in which it is straightforward to classically simulate the action of $U$ on a quantum state, for example by direct numerical integration of the equations of motion for an open quantum system.
For simplicity, we therefore treat $U$ as black-box oracle that maps a quantum state $\ket\psi$ stored in classical memory to $U\ket\psi$.
We then ask how this oracle can be used to compute an effective Pauli noise model that can be passed up the software stack to a circuit-level Clifford simulator such as \texttt{Stim} \cite{gidney2021stim} to estimate the threshold of an error-correcting code.

Our general strategy in Section \ref{sec:unitary} is to write $U$ as the target unitary $U_0$ followed by an error $U_\err$, i.e.~$U = U_\err U_0$, and expand $U_\err$ in the basis of Pauli strings.
As one might expect, we then show the Pauli coefficients of $U_\err$ provide the closest Pauli channel to $U_\err$ with respect to the distance metric induced by the Frobenius norm.
We repeat this analysis in Section \ref{sec:channel} for the case in which the approximate implementation of $U_0$ is not a unitary operator $U$, but a quantum channel $\U$.
For quick reference, our main technical results are provided in Theorem \ref{thrm:distance}, Corollary \ref{thrm:pure}, and Corollary \ref{thrm:mixed}.
Finally, we discuss some open questions and related problems, particularly relating to the treatment of coherent errors, in Section \ref{sec:discussion}.

%%%%%%%%%%%%%%%%%%%%%%%%%%%%%%%%%%%%%%%%%%%%%%%%%%
\section{An approximate unitary}
\label{sec:unitary}

Here we consider a unitary $U\approx U_0$ that approximately implements the target unitary $U_0$ on a collection of qubits.
We can generally write $U = U_\err U_0$, and note that oracle access to $U:\ket\psi\to U\ket\psi$ is straightforward to convert into oracle access to $U_\err:\ket\psi\to U_\err\ket\psi = U U_0^{-1} \ket\psi$ simply by prepending the inverse of the (known) unitary $U_0$ to the state $\ket\psi$.
We then expand
\begin{align}
  U_\err = \sum_{\text{Pauli stings}~P} u_P P,
  &&
  \text{where}
  &&
  u_P = \bk{P, U_\err} = \EE_b \bk{b|P U_\err|b}.
  \label{eq:pauli_pure}
\end{align}
Here $\bk{A, B} = \Tr(A^\dag B) / \dim(A)$ is the Frobenius inner product normalized to the dimension $\dim(A)$ of the Hilbert space addressed by $A$ (and $B$), and $\EE_b$ denotes an average over all computational basis states (bitstrings) $b\in\set{0,1}^N$.
Read from right to left, the expression $u_P = \EE_b \bk{b|P U_\err|b}$ provides a straightforward routine for computing the coefficients $u_P$:
\begin{enumerate}
  \item Prepare an initial computational basis state $\ket{b}$.
  \item Evolve this state under $P U_\err$.
  \item Project the resulting state onto $\ket{b}$.
  \item Average this projection over all bitstrings $b$.
\end{enumerate}
In practice, it may be more efficient to compute the matrix $U_\err$ once, which requires a total of $2^N$ simulations to collect the column vectors $\ket{U_\err, b} = U_\err\ket{b}$ into the matrix $\sum_b \op{U_\err, b}{b} = U_\err$, and then to compute the inner product $\bk{P,U_\err}$ for each $P$.

If the physical unitary $U$ addresses a Hilbert space that includes states outside of the computational subspace $\H_{\mathrm{comp}}$, such as the excited states of a superconducting qubit or Rydberg atom, then $U_\err$ may generally include leakage errors.In this case, the final state $U\ket\psi$ should be projected onto the computational subspace before computing the inner product with Pauli operators.
Equivalently, the Pauli $P$ can be extended have zero support outside the computational subspace.
The expression $u_P = \EE_b\bk{b|P U_\err|b}$ is then still correct as is, but the normalization of the inner product should be changed so that $\bk{A, B} = \Tr(A^\dag B)/\dim(\H_{\mathrm{comp}})$.
While the consideration of leakage error changes some technical details in this work, it does not affect our overall conclusions.
For simplicity, we therefore exclude leakage errors in the work below.

The unitary $U_\err$ is a coherent error that is \emph{not} equivalent to a Pauli channel.
Nonetheless, a semiclassical interpretation of $U_\err$ is that it applies the Pauli error $P$ with probability $\abs{u_P}^2$.
As required for their interpretation as probabilities, the numbers $\abs{u_P}^2$ are nonnegative and sum to one:
\begin{align}
  \bk{U_\err, U_\err} = \sum_{P, Q} u_P^* u_Q \bk{P, Q}
  = \sum_P \abs{u_P}^2
  = 1.
\end{align}
Here and throughout, $A^*$ is the complex conjugate of $A$.
To put our interpretation of $\abs{u_P}^2$ as the probability of having a Pauli error $P$ on stronger footing, we show that this interpretation yields the \emph{closest} Pauli noise channel to $U_\err$.
To this end, we consider the representation of $U_\err$ as a quantum channel in the form
\begin{align}
  \hat{U}_\err
  = U_\err \otimes U_\err^*
  = \sum_{P, Q} u_P u_Q^* P \otimes Q^*,
\end{align}
which by slight abuse of notation acts on a density matrix $\rho$ as $\hat{U}_\err(\rho) = U_\err \rho U_\err^\dag$.
More precisely, we can vectorize any operator $\O = \sum_{a,b} \O_{ab} \op{a}{b} \to \oket{\O} = \sum_{a,b} \O_{ab} \ket{ab}$, in which case the channel $\hat{U}_\err$ simply acts on the vectorized density matrix by matrix multiplication, taking $\oket{\rho} \to \oket{\hat{U}_\err(\rho)} = \hat{U}_\err \oket{\rho}$.
We then let $\B=\set{I_2,X,Y,Z}$ be the set of single-qubit Pauli operators, including the qubit identity operator $I_2$.
The formal justification for our derivation of an effective Pauli noise model is the following:
\begin{theorem}
  Let $\C = \sum_{P,Q\in\B^{\otimes N}} c_{PQ} P \otimes Q^*$ be an $N$-qubit quantum channel.
  The closest Pauli channel to $\C$ is $\C^\Pauli = \sum_{P\in\B^{\otimes N}} c_{PP} P \otimes P^*$, where distance is measured by the metric $d(A,B) = \norm{A-B}$ induced by the norm $\norm{A} = \sqrt{\bk{A,A}}$.
  \label{thrm:distance}
\end{theorem}
For reference, here $\bk{A, B} = \Tr(A^\dag B) / \dim(A)$ is the Frobenius inner product normalized to the dimension $\dim(A)$ of the Hilbert space addressed by $A$ and $B$.

\textit{Proof}.
Any $N$-qubit Pauli noise channel can be written in the form
\begin{align}
  \E_\Pauli = \sum_{P\in\B^{\otimes N}} e_P P \otimes P^*
\end{align}
where $e_P$ is the probability of applying the Pauli string $P$, and $\sum_P e_P = 1$.
Due to the orthogonality of the Pauli strings, the squared distance between $\C$ and $\E^\Pauli$ is
\begin{align}
  d(\C, \E_\Pauli)^2
  = \sum_P \abs{c_{PP} - e_P}^2 + \sum_{P\ne Q} \abs{c_{PQ}}^2.
\end{align}
This distance minimized by setting $e_P = c_{PP}$, which implies Theorem \ref{thrm:distance}.

As an immediate consequence of Theorem \ref{thrm:distance} and Eq.~\eqref{eq:pauli_pure}, we can state the following:
\begin{corollary}
  If $U_\err$ is an $N$-qubit unitary, the closest Pauli channel to $\hat{U}_\err = U_\err\otimes U_\err^*$ is $\hat{U}_\err^\Pauli = \sum_{P\in\B^{\otimes N}} \abs{u_P}^2 P\otimes P^*$, where $u_P = \bk{P,U_\err} = \EE_{b\in\set{0,1}^N}\bk{b|PU_\err|b}$.
  \label{thrm:pure}
\end{corollary}
As in Theorem \ref{thrm:distance}, here ``closest'' is meant with respect to the metric induced by the norm $\norm{A} = \sqrt{\bk{A,A}}$, $\B$ is the set of single-qubit Pauli operators together with the identity, and $\EE_{b\in\set{0,1}^N}$ denotes an average over all $N$-bit strings.

%%%%%%%%%%%%%%%%%%%%%%%%%%%%%%%%%%%%%%%%%%%%%%%%%%
\section{An approximate channel}
\label{sec:channel}

We now consider the calculation of an effective Pauli noise model in the context of an open quantum system, when the approximate implementation of $U_0$ is not a unitary $U$, but a channel $\U$.
For example, $\U$ may be generated by open quantum system dynamics, or obtained by averaging over unitaries $U(s)$ parametrized by random variables $s$, such that $\U = \EE_{s\sim S} U(s)\otimes U(s)^*$, where $\EE_s$ denotes an average of $s$ drawn from an appropriate ensemble $S$.
We can define the error channel
\begin{align}
  \U_\err = \U \circ \hat{U}_0^{-1} = \sum_{P,Q} w_{PQ} P \otimes Q^*,
  \label{eq:error_channel}
\end{align}
where $\hat{U}_0 = U_0 \otimes U_0^*$, and $\Phi\circ\chi$ denotes the composition of channels $\Phi$ and $\chi$, such that $(\Phi\circ\chi)(\rho) = \Phi(\chi(\rho))$.
Analogously to the unitary case in Section \ref{sec:unitary}, we seek an expression for the coefficients $w_{PP}$, which again determine the closest Pauli noise channel to $\U_\err$, namely $\U_\err^\Pauli = \sum_P w_{PP} P\otimes P^*$.
In principle, we can pick off a Pauli coefficient with
\begin{align}
  w_{PQ} = \bk{P\otimes Q^*, \U_\err},
  \label{eq:channel_coef}
\end{align}
but this expression does not tell us how to actually \emph{compute} $w_{PQ}$ using a classical oracle for $\U_\err$.
To clarify the procedure for computing $w_{PQ}$, we define the column vector $\oket{\U_\err,ab} = \U_\err\oket{ab}$ obtained by flattening $\U_\err(\op{a}{b})$ into a vector.
The matrix representation of $\U_\err$ in Eq.~\eqref{eq:error_channel} can then be obtained by computing and collecting the $4^N$ column vectors $\oket{\U_\err,ab}$ for all $a,b\in\set{0,1}^N$ into the matrix $\sum_{a,b} \oop{\U_\err,ab}{ab} = \sum_{a,b} \U_\err \oop{ab}{ab}$.
This matrix is, in fact, nearly the Choi matrix of $\U_\err$, up to array reshaping and a permutation of tensor factors.
The coefficient $w_{PQ}$ is then obtained from this matrix representation of $\U_\err$ by taking a normalized Frobenius inner product with $P\otimes Q^*$.
% As before, the state $\U_\err(\rho)$ should be projected onto the computational subspace to address leakage if necessary before computing coefficients $w_{PQ}$.

To give the expression in Eq.~\eqref{eq:channel_coef} operational meaning, we observe that for any pair of $N$-qubit channels $\Phi$ and $\chi$, the inner product $\bk{\Phi,\chi}$ is equal to the entanglement fidelity \cite{nielsen1996entanglement, nielsen2002simple} $F_\ent(\Phi^\dag\circ\chi)$, defined by
\begin{align}
  F_\ent(\C) = \bk{\phi|(\hat{I}\otimes\C)(\phi)|\phi}
  \label{eq:fidelity}
\end{align}
where $\phi = \op{\phi}$ is the density matrix of a maximally entangled state $\ket{\phi} = \frac1{\sqrt{2^N}} \sum_b \ket{bb}$ of $2N$ qubits, $\hat{I}$ is the identity channel, and $\Phi^\dag$ is the adjoint channel of $\Phi$.
Since Pauli strings are self-adjoint, the adjoint of $\Phi=\sum_{P,Q}\Phi_{PQ} P\otimes Q^*$ is $\Phi^\dag = \sum_{P,Q} \Phi_{PQ}^* P\otimes Q^*$.
Similarly expanding $\chi = \sum_{P,Q}\chi_{PQ}P\otimes Q^*$, by explicit calculation we can show that
\allowdisplaybreaks
\begin{align}
  F_\ent(\Phi^\dag\circ\chi)
  &= \frac1{4^N} \sum_{a,b,c,d} \Braket{aa|\p{\op{c}{d}\otimes(\Phi^\dag\circ\chi)(\op{c}{d})}|bb} \\
  &= \frac1{4^N} \sum_{a,b} \bk{a|(\Phi^\dag\circ\chi)(\op{a}{b})|b} \\
  &= \frac1{4^N} \sum_{a,b} \sum_{P,Q,R,S} \Phi_{PQ}^* \chi_{RS} \bk{a|PR\op{a}{b}SQ|b} \\
  &= \sum_{P,Q,R,S} \Phi_{PQ}^* \chi_{RS} \bk{P,R} \bk{S,Q} \\
  &= \bk{\Phi,\chi},
\end{align}
which implies
\begin{align}
  w_{PQ}
  = F_\ent(\sp{P\otimes Q^*}\circ \U_\err)
  = \EE_{a,b} \bk{a|P\U_\err(\op{a}{b})Q|b}.
\end{align}
As a sanity check, we can verify that this result is consistent with Eq.~\eqref{eq:pauli_pure} for computing $w_{PP} = \abs{u_P}^2$ when $\U_\err = U_\err\otimes U_\err^*$.

Altogether, we can state the following corollary to Theorem \ref{thrm:distance}:
\begin{corollary}
  If $\U_\err$ is an $N$-qubit channel, the closest Pauli channel to $\U_\err$ is $\U_\err^\Pauli = \sum_{P\in\B^{\otimes N}} w_{PP} \hat{P}$, where $w_{PP} = \bk{\hat{P},\U_\err} = F_\ent(\hat{P}\circ\U_\err)$ and $\hat{P} = P\otimes P^*$.
  \label{thrm:mixed}
\end{corollary}
Here the entanglement fidelity $F_\ent(\C)$ is defined in Eq.~\eqref{eq:fidelity}.

%%%%%%%%%%%%%%%%%%%%%%%%%%%%%%%%%%%%%%%%%%%%%%%%%%
\section{Discussion and open questions}
\label{sec:discussion}

We found the closest Pauli channel $\U_\err^\Pauli$ to a given error channel $\U_\err$, and provided procedures to compute the parameters of this Pauli channel using a classical oracle for $\U_\err$ (implemented, for example, via numerical integration of the equations of motion for an open quantum system).
Though the use of Pauli noise models is standard practice, our results leave a few open questions that warrant discussion.
We make no claim of being the first to ask these (or similar) questions, and do not claim that the answers to these questions cannot be found in existing literature; we merely claim that these are natural questions to consider in light of the technical results in Sections \ref{sec:unitary} and \ref{sec:channel}.

%%%%%%%%%%%%%%%%%%%%%%%%%%%%%%%%%%%%%%%%%%%%%%%%%%
\subsection{A better measure of approximation quality}

First, if $\U_\err$ describes a coherent error, then there is a quantitative sense in which nearly \emph{any} channel with the same fidelity as $\U_\err$ is ``nearly as good'' as $\U_\err^\Pauli$ for approximating $\U_\err$ -- as far as the Frobenius metric is concerned.
To illustrate this point, consider a unitary error channel $\hat{U}_\err = U_\err \otimes U_\err^*$ with $U_\err = \sum_P u_P P$.
Let $I$ denote the $N$-qubit identity operator and $\hat{I}=I\otimes I$ denote the $N$-qubit identity channel, and assume that $\hat{U}_\err$ is close to the identity channel, with fidelity $\bk{\hat{I},\hat{U}_\err} = \abs{\bk{I,U_\err}}^2 = \abs{u_I}^2 = 1 - \epsilon^2$ for some positive $\epsilon\ll1$.
The normalization condition $\sum_P\abs{u_P}^2=1$ implies that $\sum_{P\ne I}\abs{u_P}^2=\epsilon^2$, and in turn $\abs{u_P}=O(\epsilon)$ for all $P\ne I$.
The squared distance between $\hat{U}_\err$ and a Pauli channel $\E^\Pauli = \sum_P e_P P \otimes P^*$ is
\begin{align}
  d(\hat{U}_\err, \E^\Pauli)^2
  &= \sum_P (\abs{u_P}^2 - e_P)^2 + \sum_{P\ne Q} \abs{u_P u_Q^*}^2 \\
  &= \sum_{P\ne I} (\abs{u_P}^2 - e_P)^2 + (1 - \epsilon^2 - e_I)^2
  + \sum_{\substack{P\ne Q \\P,Q\ne I}} \abs{u_P u_Q^*}^2
  + 2 (1 - \epsilon^2) \epsilon^2.
  \label{eq:dist_identity}
\end{align}
Choosing a Pauli channel with $e_P = \abs{u_P}^2$ minimizes the distance between $\hat{U}_\err$ and $\E^\Pauli$, but this choice mostly eliminates an $O(1)$ contribution from $e_I = \abs{u_I}^2 = 1 - \epsilon^2$, and otherwise only eliminates $O(\epsilon^4)$ terms in the first sum of Eq.~\eqref{eq:dist_identity}.
In particular, the final $\sim\epsilon^2$ term in Eq.~\eqref{eq:dist_identity} is unaffected by the choice of $e_P$, so the exact choice of Pauli error channel -- as long as it has fidelity $\abs{u_P}^2$ with the identity channel -- does not seem to particularly matter as $\epsilon\to0$.

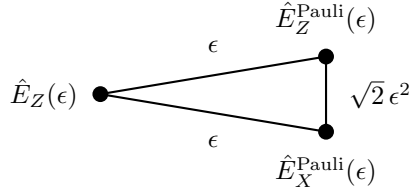
\begin{figure}
  \centering
  \begin{tikzpicture}
    \coordinate (A) at (0, 0);
    \coordinate (B) at (3, +0.5);
    \coordinate (C) at (3, -0.5);
    \draw[thick] (A) -- node[above=.5em] {$\epsilon$}
    (B) -- node[right=.5em] {$\sqrt{2}\,\epsilon^2$}
    (C) -- node[below=.5em] {$\epsilon$} (A);

    \foreach \point in {A,B,C}
    \fill (\point) circle (3pt);
    \node[left=.5em] at (A) {$\hat{E}_Z(\epsilon)$};
    \node[above=.5em] at (B) {$\hat{E}_Z^\Pauli(\epsilon)$};
    \node[below=.5em] at (C) {$\hat{E}_X^\Pauli(\epsilon)$};
  \end{tikzpicture}
  \caption{
    Geometric relationship between the error channel $\hat{E}_Z(\epsilon)$, it nearest Pauli channel approximation $\hat{E}_Z^\Pauli(\epsilon)$, and the Pauli channel $\hat{E}_X^\Pauli(\epsilon)$.
    Distances induced by the normalized Frobenius norm are indicated to second order in $\epsilon$.
    As $\epsilon\to0$, the two Pauli channels approach each other faster than either channel approaches $\hat{E}_Z(\epsilon)$.
  }
  \label{fig:triangle}
\end{figure}

As a concrete example, consider the single-qubit unitary error $E_Z(\epsilon) = e^{-\ii\epsilon Z} = \cos\epsilon - \ii \sin\epsilon Z$ with associated quantum channel
\begin{align}
  \hat{E}_Z(\epsilon) = (\cos\epsilon)^2 - \ii \cos\epsilon \sin\epsilon (I\otimes Z + Z\otimes I) - (\sin\epsilon)^2 Z \otimes Z.
\end{align}
The closest Pauli channel to $\hat{E}_Z(\epsilon)$ is
\begin{align}
  \hat{E}_Z^\Pauli(\epsilon) = (\cos\epsilon)^2 - (\sin\epsilon)^2 Z \otimes Z,
\end{align}
with squared distance
\begin{align}
  d(\hat{E}_Z(\epsilon),\hat{E}_Z^\Pauli(\epsilon))^2
  = (\cos\epsilon\sin\epsilon)^2
  = \epsilon^2  - \frac{4}{3} \epsilon^4 + O(\epsilon^6).
\end{align}
However, the Pauli channel $\hat{E}_X^\Pauli(\epsilon) = (\cos\epsilon)^2 - (\sin\epsilon)^2 X \otimes X$ (for example) is just as close to $\hat{E}_Z(\epsilon)$ at leading order in $\epsilon$:
\begin{align}
  d(\hat{E}_X^\Pauli(\epsilon),\hat{E}_Z(\epsilon))^2
  = (\cos\epsilon\sin\epsilon)^2 + 2 (\sin\epsilon)^4
  = \epsilon^2  + \frac{2}{3} \epsilon^4 + O(\epsilon^6).
\end{align}
In this sense, $\hat{E}_X^\Pauli(\epsilon)$ is nearly as good of a choice as $\hat{E}_Z^\Pauli(\epsilon)$ to approximate $\hat{E}_Z(\epsilon)$ when $\epsilon\ll1$.
Moreover, these two Pauli channels are much closer to each other than they are to the unitary error channel, since
\begin{align}
  d(\hat{E}_X^\Pauli(\epsilon),\hat{E}_Z^\Pauli(\epsilon))^2
  = 2 (\sin\epsilon)^4 = 2 \epsilon^4 + O(\epsilon^6).
\end{align}
The geometric relationship between these three channels to second order in $\epsilon$ is sketched out in Figure \ref{fig:triangle}.

In some ways, the similarity between $\hat{E}_Z^\Pauli(\epsilon)$ and $\hat{E}_X^\Pauli(\epsilon)$ is not surprising.
With high probability, $(\cos\epsilon)^2 = 1 - \epsilon^2 + O(\epsilon^4)$, applying either of these channels to the quantum state $\rho$ returns back the same state $\rho$.
However, there is clearly a qualitative difference between $\hat{E}_Z^\Pauli(\epsilon)$ and $\hat{E}_X^\Pauli(\epsilon)$ that becomes strikingly important when considering errors in a recovery channel for a QEC code: the the $Z$-type Pauli channel $\hat{E}_Z^\Pauli(\epsilon)$ is expected to reasonably model the unitary error $\hat{E}_Z(\epsilon)$, whereas the $X$-type Pauli channel $\hat{E}_X^\Pauli(\epsilon)$ is not.
If the errors in a quantum computer have a significant $Z$-bias, for example, this bias can be leveraged to construct QEC codes with significantly higher error thresholds \cite{bonillaataides2021xzzx}.
In the limit of infinite bias, this threshold can approach 50\% (in contrast to thresholds of $\sim10^{-3}$ for the surface code).
The high threshold of a bias-tailored QEC code in a $Z$-biased quantum computing platform can be reproduced with $Z$-type Pauli error channels, but not with their $X$-type counterparts.
This observation begs the question: how can the distinction between $\hat{E}_Z^\Pauli(\epsilon)$ and $\hat{E}_X^\Pauli(\epsilon)$ -- at least for the purposes of approximating $\hat{E}_Z(\epsilon)$ in the context of QEC -- be captured in a quantitative manner that does not appear to vanish at leading order in $\epsilon$?

%%%%%%%%%%%%%%%%%%%%%%%%%%%%%%%%%%%%%%%%%%%%%%%%%%
\subsection{A better treatment of coherent errors}

Second, classically simulating Clifford circuits with Pauli channels is efficient.
For this reason, it may be reasonable (and is certainly commonplace) to replace a unitary errors by Pauli channels in Clifford circuits, such as the syndrome extraction circuit of qubit-QEC code.
As discussed above, however, taking care to choose the closest Pauli channel only addresses sub-leading discrepancies between the true error $\U_\err$ and its model $\U_\err^\Pauli$ (as measured by the Frobenius distance).
It is therefore natural to ask: is there any way to mitigate the \textit{leading-order} discrepancies?
These discrepancies arise from the off-diagonal components of the coefficient matrix $w_{PQ}$ in the expansion $\U_\err = \sum_{P,Q} w_{PQ} P\otimes Q^*$.
An efficient algorithm to address the leading-order discrepancies is clearly not possible in general, since efficiently accounting for unitary errors would amount to a classical simulation of (potentially arbitrary) quantum dynamics.
However, one may wonder (for example) whether there is a systematic approximation scheme that becomes tractable as $\epsilon\to0$.

As a post-hoc motivation for this problem, we note that there is a tension in the literature regarding coherent errors in the context of QEC.
Ref.~\cite{beale2018quantum} proves that QEC -- and syndrome extraction in particular -- generally decoheres unitary errors, making their effect converge to that of stochastic (Pauli) noise.
Ref.~\cite{huang2019performance} also found that QEC suppresses coherent errors much \textit{more} favorably than Pauli errors: with a unitary error model, the diamond norm error $D_\diamond'$ of a logical qubit (after a single faulty QEC code cycle) is bounded by the diamond norm error $D_\diamond$ of a physical qubit (after a faulty single gate) as $D_\diamond'=O(D_\diamond^d)$ for coherent errors in a distance-$d$ code.
For a Pauli noise model, $D_\diamond' = O(D_\diamond^{(d+1)/2})$.
However, when examining concatenated codes (with $n$ concatenation levels of distance-$d$ codes), Ref.~\cite{greenbaum2018modeling} found that coherent errors are negligible only up to some number ($\sim1/\epsilon^{d^n-1}$) of error correction cycles, after which coherent errors induce higher rates of logical failure than a Pauli noise model would predict.
Numerical studies \cite{bravyi2018correcting} also provide evidence that that approximating coherent errors with Pauli noise models can yield an accurate estimate of a QEC code threshold, but underestimates the logical error rate of a code in the sub-threshold regime.
Taken together, these results paint a murky picture for the proper treatment and expected severity of coherent errors in QEC, which motivates the need for improved modeling techniques.
In an era of rapid hardware development for quantum computing devices, there is also an imminent need to better model and understand the effects of unitary errors for quantum device characterization, verification, and validation (QCVV).

As a final point, we note that there have been some proposals to circumvent the problem of coherent errors by using classically randomized compiling techniques, such as Pauli or Clifford twirling, that ``decohere'' unitary errors \cite{kern2005quantum, geller2013efficient, wallman2016noise, cai2019constructing}.
However, these techniques generally come at the cost of averaging out structure in circuit errors (for example, by converting $Z$-biased errors into depolarizing errors), which could otherwise be leveraged to design higher-threshold QEC codes \cite{bonillaataides2021xzzx, xu2023tailored, roffe2022biastailored}.
It may therefore be desirable to address coherent errors directly, rather than converting them into stochastic Pauli errors.

\bibliography{main.bib}

\end{document}